\documentclass[conference]{IEEEtran}
\IEEEoverridecommandlockouts

\usepackage{cite}                   
\usepackage{amsmath, amssymb, amsfonts} 
\usepackage{algorithmic}            
\usepackage{graphicx}               
\usepackage{textcomp}               
\usepackage{xcolor}                 
\usepackage{censor}
\usepackage{multirow}               
\usepackage{tikz}                   
\usetikzlibrary{matrix}             
\usepackage{booktabs}               
\usepackage{array}                  
\usepackage{geometry}               
\usepackage{longtable}              
\usepackage{pdflscape}              
\usepackage{rotating}               
\usepackage{tabularx}               
\usepackage{float}                  
\usepackage{xspace}                 

\usepackage{babel}                  
\usepackage{mdframed}               
\usepackage{etoolbox}               

\usepackage{tabularray}
\usepackage{caption}

\usepackage{tikz}
\usepackage{xcolor}
\usetikzlibrary{shapes,arrows,positioning,fit,backgrounds,calc}

\definecolor{iteration1}{RGB}{200,220,240}
\definecolor{iteration2}{RGB}{160,200,230}
\definecolor{iteration3}{RGB}{120,180,220}
\definecolor{arrowcolor}{RGB}{80,120,180}
\definecolor{textcolor}{RGB}{50,50,50}

\def\numberdimensions{22}
\def\numbercharacteristics{113}
\def\numberiterations{three}

\definecolor{ieeeblue}{RGB}{0,120,177} 
\usepackage[colorlinks = true,        
            linkcolor= ieeeblue,
            urlcolor  = ieeeblue,
            citecolor = ieeeblue,
            anchorcolor = ieeeblue]{hyperref}

\newcommand{\etal}{~\textit{et~al.}~}  

\def\BibTeX{{\rm B\kern-.05em{\sc i\kern-.025em b}\kern-.08em
    T\kern-.1667em\lower.7ex\hbox{E}\kern-.125emX}}

\begin{document}

\title{SoK: A Taxonomy for Distributed-Ledger-Based \\Identity Management}
\author{\IEEEauthorblockN{Awid Vaziry\IEEEauthorrefmark{1},
Sandro Rodriguez Garzon\IEEEauthorrefmark{2}, Patrick Herbke\IEEEauthorrefmark{3}, Carlo Segat\IEEEauthorrefmark{4} and
Axel Küpper\IEEEauthorrefmark{5}}
\IEEEauthorblockA{Service-centric Networking,
Technische Universität Berlin\\
Berlin, Germany \\
Email: \IEEEauthorrefmark{1}vaziry@tu-berlin.de,
\IEEEauthorrefmark{2}sandro.rodriguezgarzon@tu-berlin.de,\\
\IEEEauthorrefmark{3}p.herbke@tu-berlin.de,
\IEEEauthorrefmark{4}carlo.segat@tu-berlin.de,
\IEEEauthorrefmark{5}axel.kuepper@tu-berlin.de}}


\author{\IEEEauthorblockN{Awid Vaziry, Sandro Rodriguez Garzon, Patrick Herbke, Carlo Segat, Axel Küpper}
\IEEEauthorblockA{Service-centric Networking \\
Technische Universit\"at Berlin /  T-Labs\\
Berlin, Germany \\
\{vaziry\}\textbar \{sandro.rodriguezgarzon\}\textbar \{p.herbke\}\textbar \{carlo.segat\}\textbar \{axel.kuepper\}@tu-berlin.de}}

\maketitle

\begin{abstract}
The intersection of blockchain (distributed ledger) and identity management lacks a comprehensive architectural framework for classifying distributed-ledger-based identity solutions. This paper presents a methodologically developed taxonomy from analyzing 390 scientific papers and expert discussions. The resulting artifact comprises \numberdimensions\ dimensions with \numbercharacteristics\ characteristics organized into three groups: trust anchor implementations, identity architectures (identifiers and credentials), and ledger specifications. The taxonomy enables systematic analysis, comparison, and design of distributed-ledger-based identity solutions, as demonstrated through application to two distinct architectures. As the first methodology-driven taxonomy in this domain, this work advances standardization and enhances understanding of distributed-ledger-based identity architectures, providing researchers and practitioners with a structured framework for evaluating design decisions and implementation approaches.
\end{abstract}



\section{Introduction}
\label{sec:introduction}

\IEEEPARstart{U}{tilizing} blockchains and distributed ledgers for digital identities in future identity management (IDM) systems promises better trust, security, user control over personal data\cite{dunphy_first_2018}. Yet, limited standardization, interoperability, and poor user experience of current solutions continue to hinder the widespread adoption of secure ledger-based identity systems~\cite{liu_blockchain-based_2020}. Various research efforts, such as those by \textit{Schardong}\etal~\cite{schardong_self-sovereign_2022} and \textit{Lesavre}\etal~\cite{lesavre_taxonomic_2020} aim to conceptualize the current landscape of blockchain-enabled IDM solutions. However, from an architectural point of view, there is a lack of a holistic and unified conceptual framework to characterize and assess those solutions. This impedes effective communication and collaboration among researchers and application developers. To improve the organization of this field and to promote greater consistency and standardization, it is essential to organize and structure knowledge in the area of distributed-ledger-based IDM solutions with a focus on architectural design.

In this paper, we present the first comprehensive taxonomy for system architectures of distributed-ledger-based identity solutions, systematically organizing relevant characteristics to guide researchers and practitioners. Inspired by \textit{Glass}\etal\cite{glass_contemporary_1995}, a taxonomy is defined as a system of groupings that organizes and categorizes knowledge of a field, enabling researchers to study relationships between concepts, hypothesize about new ideas, and evaluate existing architectures. We employ a rigorous, iterative approach following the established guidelines of \textit{Nickerson}\etal \cite{nickerson_method_2013} and \textit{Kundisch}\etal\cite{kundisch_update_2022}. Our taxonomy emerges from an extensive analysis of 390 scientific papers from the field and insights from multiple expert discussion panels, with a special focus on the architectural aspects of IDM solutions leveraging Distributed Ledger Technology (DLT). The resulting artifact comprises \numberdimensions\ dimensions with \numbercharacteristics\ characteristics, refined through \numberiterations\ methodical iterations. To validate the taxonomy's practical utility and analytical power, we apply it exemplarily to two representative system architectures. It effectively classifies the solutions, highlighting key design decisions and potential shortcomings that might otherwise remain overlooked.

The article starts with a discussion of blockchain and IDM fundamentals in Section ~\ref{sec:background}, followed by a review of relevant taxonomies in Section~\ref{sec:related-work}. The taxonomy's development methodology is presented in Section~\ref{sec:method}, while the taxonomy itself and its evaluation are detailed in Sections~\ref{sec:taxonomy-blockchain-identity}~and~\ref{sec:evaluation}. Limitations and conclusions are discussed in Sections~\ref{sec:discussion}~and~\ref{sec:conclusion}.               
\section{Background}
\label{sec:background}

This section discusses the technical and theoretical concepts of blockchain, digital identity. It explores key identity-related terminology, decentralized identity frameworks, and DLTs, providing a foundation for this work

\subsection{Identity Concepts and Terminology}
Identity is defined by the International Organization for Standardization (ISO) as a "\textit{set of attributes related to an entity}"\cite{iso_24760-1:2019}, where an entity may be a person, organization, server, application, or service\cite{ITU-T_X.1252}. Building on this, we define identity as a collection of characteristic attributes that uniquely describe a physical or non-physical being within a specific system. Digital identity extends this concept to digital realm. Inspired by the definitions given by NIST~\cite{grassi_digitalID_800-63-3_2017}, Domingo et al.\cite{domingo_how_2020}, and Sedlmeir et al.\cite{sedlmeir_digital_2021}, we define digital identity as a digital collection of attributes uniquely describing a physical or non-physical being within a specific system, enabling participation within this specific system and responding to identity-related transactions. IDM encompasses then the processes and technologies to create, manage, and authenticate identities. Key elements include identifiers (unique labels for digital identities), authentication (verification of identity control), authenticators (verification tools), and authorization (managing access)~\cite{ITU-T_X.1252, grassi_digitalID_800-63-3_2017}.



\subsection{Decentralized Digital Identity}

User-centric identity systems have gained momentum in recent years \cite{yan_blockchain-driven_2024,liu_blockchain-based_2020}, driven by the privacy-preserving paradigm of Self-Sovereign Identity (SSI), alongside the technical concepts of Decentralized Identifiers (DIDs) \cite{sporny_decentralized_2022}, and Verifiable Credentials (VCs) \cite{sporny_verifiable_2022}. SSI proclaims a system design that empowers individuals to control their digital identities without relying on central authorities by leveraging distributed ledgers for secure verification~\cite{dunphy_first_2018}. DIDs provide unique identification in a decentralized manner, resolving to DID documents containing metadata, cryptographic keys, and service endpoints. The DID documents of public DIDs are stored in verifiable data registries such as distributed ledgers, while their counterparts of private DIDs are derived from the identifier. VCs contain claims about a subject, must include the subject's identifier (often a DID), and are signed by an issuer to ensure integrity and non-repudiation. Holders can generate Verifiable Presentations (VPs) from VCs to selectively share verified information~\cite{sporny_verifiable_2022, grassi_digitalID_800-63-3_2017}.


\subsection{Identity Roles and Trust Relationships}

The issuer-holder-verifier model is fundamental to all decentralized identity frameworks. Issuers create and sign VCs about subjects. Holders (often the subjects themselves) control these VCs and determine when and with whom to share them through VPs. Verifiers authenticate the presented VPs, confirming their validity and provenance~\cite{sporny_verifiable_2022}. Traditional federated identity models operate with different roles. Identity Providers (IdPs) manage identity information and issue assertions about users. Relying Parties (RPs) or service providers trust these IdPs to verify user identities and validate assertions~\cite{windley_learning_2023}. These established trust relationships form the foundation of identity transactions across both traditional and decentralized systems. Central to these are trust anchors which are fundamental elements that serve as the basis to establish trust in identity systems. Trust anchors enable verification of credentials and assertions without requiring direct trust relationships between all entities. In traditional PKI systems, certificate authorities typically serve as trust anchors. In distributed ledger-based identity systems, trust anchors may include the distributed ledger itself, governance frameworks, or cryptographic keys registered on the ledger~\cite{sporny_decentralized_2022, windley_learning_2023}.


\subsection{Distributed Ledger Technology}

DLT enables the maintenance of append-only transactional databases in a decentralized manner. Blockchain, the most prominent DLT implementation, secures data by bundling transactions into cryptographically linked blocks. This concept was introduced for the Bitcoin cryptocurrency~\cite{nakamoto_bitcoin_2008}, while Ethereum extended it with smart contracts for programmable transactions~\cite{buterin_next-generation_2014}. DLT networks are categorized by their permission models. Permissioned networks restrict who can read, write, and host nodes, typically governed by a single entity or a consortium~\cite{androulaki_hyperledger_2018}. Permissionless networks allow anyone to participate but they often require transaction fees to prevent spam and to compensate node operators. These fees, denominated in units like "gas" in Ethereum, create resource constraints that impact performance~\cite{palma_what_2021}. For IDM, DLT provides promising capabilities including immutable record-keeping and secure sharing of verification material, without a single point of failure. These properties address fundamental challenges in traditional IDM systems, particularly regarding security, privacy, and user control.


\section{Related Work}
\label{sec:related-work}

Various taxonomies have addressed DLT in combination with IDM in academic literature. \textit{Lesavre et al.}\cite{lesavre_taxonomic_2020} categorized DLT-based IDM systems based on architectures and governance models, emphasizing terminology and emerging standards. Yet, their work lacks a taxonomy with explicit dimensions and characteristics. \textit{Schardong et al.}\cite{schardong_self-sovereign_2022} conducted a structured literature review of over 80 articles on SSI, creating a hierarchical taxonomy. While contributing significantly to SSI research, their work did not explicitly address DLT-based solutions and used a tree-like structure rather than a dimensional framework. \textit{Ngo et al.}\cite{ngo_systematic_2023} analyzed 361 articles on IDM with blockchain, focusing on research trends and metadata rather than architectural classification. \textit{Amard et al.}\cite{amard_designing_2024} developed a taxonomy examining governance choices in digital identity infrastructures, using Nickerson's methodology but with a different analytical lens than our architectural design focus. \textit{Yan}\etal~\cite{yan_blockchain-driven_2024} conducted a comprehensive interdisciplinary review of decentralized IDM, analyzing 149 articles to develop a "Task Structure-Technological Properties-Fit" framework. Their work identified two key task goals (identity value creation and maintenance), three stakeholder levels, and two technological properties (interoperability and self-sovereignty). While their framework provides valuable insights into the alignment between tasks and technologies, it focuses on the application of decentralized IDM across contexts rather than an architectural classification.

Existing taxonomies address specific sub-fields of DLT-based IDM, but their scope remains insufficient for categorizing contemporary solutions. Our research bridges this gap through a comprehensive methodology combining extensive literature review with expert consultations. The resulting framework enables systematic classification, analysis, and evaluation of DLT-based IDM architectures, advancing both theoretical understanding and practical implementation in this domain.                 
\section{Methodology}
\label{sec:method}

A taxonomy should be concise yet robust, comprehensive, expandable, and explanatory~\cite{nickerson_method_2013, kundisch_update_2022}. Conciseness requires limited dimensions and characteristics, while robustness ensures sufficient differentiation of the objects of interest. Comprehensiveness demands classification of all known objects within the domain, and extendibility allows for the inclusion of new dimensions as the field evolves. The explanatory nature provides clear descriptions that enhance understanding rather than merely listing features. The taxonomy development process presented here follows an iterative approach guided by a meta-characteristic that defines the taxonomy's focus and scope.

\subsection{Problem and Objectives}

In DLT-based IDM, solutions range from fully on-chain systems to those with minimal blockchain integration. However, there is no structured method to assess and compare these solutions, particularly regarding critical aspects such as trust sources, governance frameworks, and credential management. Researchers and practitioners developing these solutions often fail to report key dimensions of the architecture, hindering comprehensive evaluation and comparison. This taxonomy addresses this gap by providing a classification framework for researchers and practitioners, enabling effective communication and comparison of system architectures.

\subsection{Meta-Characteristic and Approach}
\label{sec:meta-characteristic}

The meta-characteristic serves as the fundamental lens through which the phenomenon is observed and classified, guiding the selection of all dimensions and characteristics in the taxonomy. After careful consideration of the research objectives and target users, the meta-characteristic is defined as: \textbf{System Architectures of IDM Solutions utilizing Distributed Ledger Technology}. 

\subsection{Ending Conditions and Evaluation Goals}

The taxonomy development follows Nickerson's methodology~\cite{nickerson_method_2013}, employing eight objective conditions (covering completeness, consistency, uniqueness, and utility) and five subjective conditions (addressing conciseness, robustness, comprehensiveness, extendibility, and explanatory power). These conditions determine when the iterative development concludes. The taxonomy aims to provide a comprehensive framework spanning the complete spectrum of DLT-based IDM solutions while facilitating conceptual comparability through systematic classification of fundamental architectural components.

The evaluation criteria for the taxonomy are defined as: \textit{Provision of a comprehensive taxonomy encompassing the complete spectrum of DLT-based IDM solutions} and \textit{Facilitation of conceptual comparability through the identification and classification of fundamental architectural components}.

\section{Taxonomy Development}
\label{sec:taxonomy-blockchain-identity}

\begin{figure*}[t!]
\centerline{\includegraphics[scale=0.8]{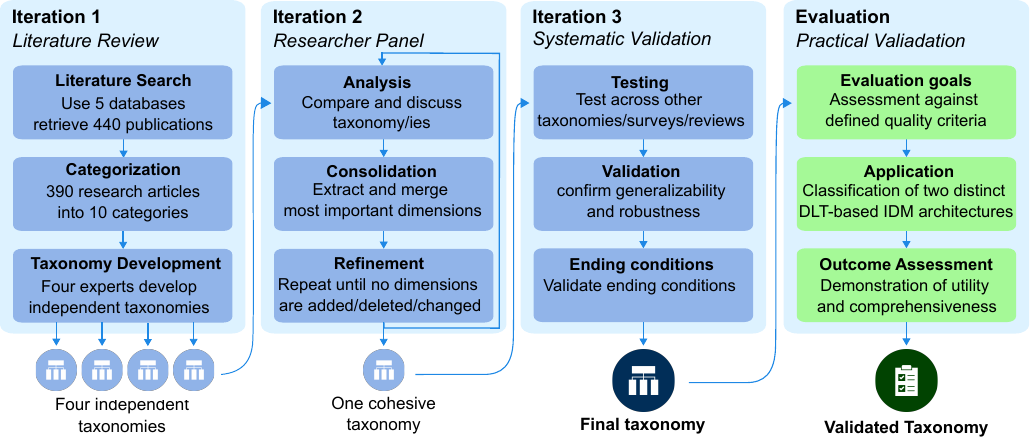}}
\caption{Taxonomy Development and Evaluation Process.}
\label{fig:taxonomy-development}
\end{figure*}

This section describes the iterative development process for creating the taxonomy. The approach was structured and included a literature review, discussions with an expert panel, and validation against established taxonomies. \autoref{fig:taxonomy-development} illustrates the development and evaluation process.

\subsection{Literature-Based Development (Iteration 1)}
The first iteration followed an empirical-to-conceptual approach through a structured literature review. The search used the terms "blockchain" OR "distributed ledger" AND "identity" in publication titles across five databases (IEEE Xplore, SpringerLink, ACM Digital Library, Wiley Online Library, Science Direct) in October 2024. After removing duplicates, 440 articles were identified.

A categorization process was applied to organize the literature into coherent groups, resulting in the classification shown in \autoref{tab:paper_categorization}. The 51 articles in the "surveys and reviews" category were set aside for later analysis, while the remaining 390 articles were distributed among four experts for review. Each expert developed their own taxonomy based on their share of articles, guided by the meta-characteristic defined in Section \ref{sec:meta-characteristic}.

\subsection{Expert Panel Integration (Iteration 2)}
The second iteration used an empirical-to-conceptual approach to integrate the four independently developed taxonomies. A panel of researchers analyzed similarities, differences, and limitations across these taxonomies, extracting the most important dimensions and characteristics. Through multiple discussions, the panel refined and consolidated these elements into a single, cohesive taxonomy.

\subsection{Validation and Refinement (Iteration 3)}
The third iteration combined conceptual-to-empirical and empirical-to-conceptual methodologies to validate the taxonomy. This phase leveraged existing surveys, reviews, and taxonomies from subfields within DLT and identity management. The taxonomy was systematically tested against prior research to confirm its generalizability and robustness across diverse solution architectures. This iteration concluded the development process with all defined ending conditions met.

\begin{table}[!htbp]
\centering
\caption{Articles reviewed by experts, sorted into technology categories and ordered by count.}
\label{tab:paper_categorization}
\setlength{\tabcolsep}{8pt} 
\renewcommand{\arraystretch}{1.5} 
\begin{tabular}{lcl}
\hline
\textbf{Technology Categories} & \textbf{Count} & \textbf{References}  \\
\hline
SSI: Applications \& Use Cases & 69 & \cite{naik_self-sovereign_2020, takemiya_sora_2018, s_decentralized_2022, goswami_self-sovereign_2024, chen_blockchain-empowered_2022, naik_uport_2020, bandara_blockchain_2021, terzi_decentralizing_2021, jamal_blockchain-based_2019, bandara_blockchain_2021-1, dong_blockchain-aided_2021, tonu_block_2019, mansoori_hei-bct_2022, dima_scholarium_2018, singh_sidhu_trust_2022, rede_blockchain_2023, saragih_use_2022, kim_service_2023, cali_improved_2022, sreenath_blockchain_2024, thorve_decentralized_2022, kumar_administration_2022, soltani_new_2018, xu_identity_2020, stokkink_deployment_2018, datta_secured_2020, terzi_securing_2020, norta_blockchain_2022, zeydan_blockchain-based_2023, popa_chaindiscipline_2023, chen_self-sovereign_2021, alessandria_self-sovereign_2021, rahat_blockchain_2022, sahi_self-sovereign_2023, ghosh_blockchain-based_2024, santhosh_krishna_self_2023, abid_blockchain-based_2022, zeydan_blockchain-based_2024, de_salve_algoid_2023, mebrahtom_trust_2023, li_blockchain-based_2024, zeydan_blockchain-based_2023-1, becirovic_blockchain_2022, tonoy_unlocking_2024, zeydan_data_2024, otta_identity_2022, zhang_croauth_2024, barolli_sovereignrx_2024, kaiser_healthcare_2023, arai_promize_2021, farao_inchain_2024, ye_connect_2021, kumar_blockchain-based_2025, karrupusamy_design_2020, visvizi_corporate_2023, khanna_self-sovereign_2022, arai_developing_2023, patnaik_feasibility_2021, kopke_chorssi_2023, elbiaze_novel_2021, arai_novel_2020, norta_real-world_2024, seidel_dezentrale_2023, ishmaev_sovereignty_2021, shuaib_self-sovereign_2023, zhuang_self-sovereign_2023, stockburger_blockchain-enabled_2021, shuaib_self-sovereign_2022, kreps_self-sovereign_2020} \\
Authentication \& Access Control & 66 & \cite{valentin_vallois_blockchain-based_2021, ao_secure_2019, decusatis_identity-based_2018, chowdhary_blockchain_2021, zhao_design_2020, gururaj_identity_2020, mukhandi_blockchain-based_2022, huang_power_2020, biswas_secure_2021, zhou_research_2020, zhu_decentralized_2021, chakravarty_blockchain-enhanced_2018, dang_multi-dimensional_2020, cao_design_2022, drga_detecting_2023, aanandaram_blockchain-based_2024, xu_blockchain-based_2024, cui_hybrid_2020, malik_blockchain_2018, mikula_identity_2018, xiang_permissioned_2020, george_secure_2020, hegde_ddmia_2022, varshney_blockchain-based_2022, ismail_towards_2022, xiong_bdim_2023, lv_highly_2022, dissanayake_trust_2021, das_bsciam_2024, chen_new_2024, bao_smart-pki_2023, chen_lightweight_2023, joshua_design_2024, song_blockchain_2023, lian_blockchain_2023, sunitha_user_2024, xin_t-atmchain_2024, chen_research_2022, dong_blockchain-based_2023, treiblmaier_blockchain_2019, de_identity_2022, panda_cloud_2021, chen_research_2020, huang_blockchain_2021, joshi_decentralized_2019, liu_trusted_2021, han_secure_2020, bao_two-factor_2023, qian_certificateless_2022, yang_use_2024, gupta_iot_2023, you_blockchain-based_2022, fu_non-transferable_2023, wen_open_2020, wang_distributed_2017, leiding2017mapping, huang_service_2020, kara_voipchain_2023, liao_blockchain-based_2022, tian_feasibility_2020, wang2024ebias, gao_privacy-preserving_2021, wang_novel_2020, zhao_novel_2022, babu_distributed_2022, ganeshan_secure_2023}
\\
Security, Privacy, \& Trust & 47 & \cite{padmanegara_blockchain_2023, priya_efficient_2020, guo_wischain_2018, sani_idenx_2020, li_hierarchical_2020, zeng_persistent_2021, v_single_2022, costales_impact_2023, sharma_ai_2024, zhou_trustful_2024, goyal_iot_2024, kravitz_securing_2017, le_biff_2018, bendiab_wip_2018, bhattacharya_enhancing_2020, deb_covchain_2021, zheng_blockchain-based_2019, ren_bia_2020, zheng_privacy-aware_2023, arafat_blosecr_2022, xu_dynamic_2023, alamri_cybersecuriy_trust_2024, pramanik_blockchainbased_2022, sakka_blockchain_2024, janani_security_2024, mendling_achieving_2018, rathee_secure_2021, suresh_identity_2021, salem_blockchain-based_2024, yuan_identity-based_2022, garg_decentralized_2023, prieto_imbua_2020, liu_credible_2021, zhuang_bcppt_2022, dai_application_2021, abraham_protection_2022, prieto_attacks_2022, pan_collusion_2020, sun_blockchain-based_2022, xue_identity-based_2019, li_advanced_2023, paterson_balancing_2021, norta_safeguarding_2019, wang_blockchain_2024, awasthi_rfdb_2024, sharma_blockchainbased_2022, janani_span_2024}
\\
Decentralized Identifiers (DIDs) & 44 & \cite{dixit_decentralized_2022,amin_identity--things_2020,dabrowski_blockchain-based_2022,mudliar_comprehensive_2018,chen_bidm_2021,malik_blockchain_2019,liang_identity_2019,htet_blockchain_2020,zhang_blockchain_2022,srivastava_decentralization_2023,nehra_decentralized_2024,katta_blockchain-based_2023,peng_vdid_2024,omar_identity_2018,asamoah_zero-chain_2020,maldonado-ruiz_innovative_2021,ghosh_decentralized_2021,liu_ss-did_2024,belurgikar_identity_2019,chen_callchain_2021,zou_application_2022,zhan_distributed_2021,hao_tdid_2023,ngoran_blockchain-based_2023,varfolomeev_blockchain_2023,ozturk_blockchain_2023,guo_decentralized_2023,dai_smart_2021,sharma_when_2024,li_dsbt_2024,wang_next_2021,sharma_personal_2020,yu_updatable_2023,hatti_blockchain-based_2024,prieto_digital_2023,barolli_cross_2024,svetinovic_secure_2022,das_secure_2023,zeydan_blockchain-based-DID_2024,gong_toward_2024,tyagi_blockchainbased_2021,li_establishing_2022,gupta_blockchainbased_2024,garg_identity_2022}
\\
SSI: Technical Architectures & 39 & \cite{torres_tutorial_2021, naik_sovrin_2021, bhattacharjee_digiblock_2022, zeydan_blockchain-based-SSI_2024, liu_design_2020, song_digital_2022, kirupanithi_self-sovereign_chameleon_2021, srivastava_secure_2023, emati_feasibility_2023, ferdous_search_2019, liu_identity_2017, malik_tradechain_2021, gulati_self-sovereign_2019, mishra_pseudo-biometric_2021, ahmed_self-sovereign_2020, bakare_blockchain_2021, imtiaz_smart_2023, zeydan_enhanced_2024, kirupanithi_hierarchical_2021, zeydan_post-quantum_2024, sekar_decentralized_2024, k_sign_2024, al_badi_enhancing_2023, moriyama_permissionless_2022, zeydan_integrating_2024, garcia-alfaro_who_2017, kacprzyk_blockchain-based_2023, tuba_changing_2023, marrella_matching_2022, kaili_disposable_2021, chen_bbm_2020, su_blockchain-based_2022, tian_towards_2022, tamane_blockchain_2021, de_salve_multi-layer_2023, ferdous_leveraging_2023, agarkar_blockchain_2024, seifert_digital_2020, schlatt_designing_2022}
\\
Identity Governance & 38 & \cite{naik_governing_2020, mecozzi_blockchain-related_2022, al-musawi_transforming_2020, choudhari_interoperable_2021, zhao_research_2021, qian_regulated_2023, habib_blockchain_2023, lee_study_2022, grabatin_reliability_2018, gruner_quantifiable_2018, theodouli_towards_2020, furfaro_infrastructure_2019, sabir_practical_2020, lemieux_addressing_2021, banerjee_identity_2021, alom_dynamic_2021, tao_structural_2023, peralta-velecela_digital_2021, liu_public_2021, kumar_mohanta_identity_2022, zhang_privacy-preserving_2024, li_decentrlized_2021, hong_acbf_2024, gupta_digital_2021, shobanadevi_novel_2022, tavares_utilizing_2024, smys_none_2023, el_haddouti_fedidchain_2023, panetto_integrating_2018, shuhan_decentralised_2024, di_ciccio_towards_2019, hansen_blockchain-based_2018, furnell_towards_2018, si_identity-based_2020, shao_attrichain_2020, li_efficient_2021, yawalkar_integrated_2023, magar_blockchain-based_2023}
 \\
Cryptographic Algorithms & 30 & \cite{xiong_blockchain-enabled_2024, anwar_framework_2019, maldonado-ruiz_3bi-ecc_2020, zhang_identity_2021, gupta_decentralized_2023, yu_blockchain-based_2023, duan_iam-bdss_2022, archana_enhanced_2024, lin_new_2018, yang_efficient_2022, zhou_authentication_2018, srivastava_blockchain-envisioned_2022, wan_hibechain_2023, zhang_identity-based_2022, cai_cooperative_2024, wang_sharing_2022, geetha_advanced_2024, li_efficient_2024, bao_identity-based_2023, alotaibi_blockchain_2023, zheng_blockchain_2020, stawicki_digital_2023, barolli_design_2022, yu_identity-based_2024, yu_identity-based_2023, bianchi_verification_2024, han_identity-based_2020, careja_digital_2023, zhang_cross-domain_2023, wang_analysis_2023}
\\
Identity Anonymization & 25 & \cite{kirupanithi_self-sovereign_2021, zhao_blockchain_2019, khan_blockchain_2020, kim_graph_2022, rattanabunno_eoa_2023, xia_efficient_2023, raju_identity_2017, liu_graph_2022, bao_pbidm_2023, gong-guo_personal_2021, zhai_fine-grained_2022, liang_ipp-hf_2022, li_trading_2023, huang_commitment_2023, zhang_stateless_2024, prabha_rshealth_2024, li_blockchain_2019, zhang_survey_2024, dai_identity_2021, lai_bipp_2022, devidas_identity_2023, li_physical_2022, sarier_efficient_2021, yanhui_research_2022, wang_congradetect_2021}
\\
Zero-Knowledge Proofs & 21 & \cite{raipurkar_digital_2023, augot_transforming_2017, heo_decentralised_2024, yin_smartdid_2023, li_blockchain-enabled_2020, luong_zkp_privacy-preserving_2023, dieye_self-sovereign_2023, akram_case_2022, jose_diaz_rivera_securing_2024, jiang_unified_2020, tang_grac_2024, song_blockchain-based_2024, dao_securing_2023, jin_blockchain-enhanced_2024, maimut_using_2021, wang_hybrid_2023, sarier_comments_2021, tian_identity-based_2024, lourinho_securing_2021, yang_zero-knowledge-proof-based_2020, prada-delgado_puf-derived_2020}
\\
Cloud and Edge Computing & 6 & \cite{varfolomeev_blockchain_cloud_2023, ahmad_improving_2018, sathio_pervasive_2021, xu_fog-enabled_2023, pavithran_edge-based_2021, mu_identity_2022}
\\
No/Other Technology & 4 & \cite{uteyev_development_2024, haber_blockchain_2020, r_digital_2024, islam_correction_2021}
\\
\hline
\end{tabular}
\end{table}
\section{Taxonomy artifact}
The final taxonomy artifact (\autoref{tab:taxonomy-v2-sandro}) comprises \numberdimensions\ dimensions and \numbercharacteristics\ serving as a guide for understanding and classifying DLT-based identity management solutions. It highlights the complexity of approaches, assisting in identifying the strengths and trade-offs of each method. The following is a description of each dimension and selected corresponding characteristics. We categorized every dimension into one of three main groups: 1) Trust Anchor, 2) Identity, and 3) Ledger.

\textbf{Dimension Group 1: Trust Anchor:}
In this taxonomy, the dimensions of trust anchors are examined with an emphasis on the technical implementation of trusted components, such as how trust is established in verification materials (e.g., public keys), credential status information, and identifier-to-subject bindings. These technical mechanisms form the foundation upon which higher-level trust in issuers, credentials, and identity claims is built.

\subsubsection{Trust Anchor Purpose}
Each trust anchor has a specific role based on the involved entities and activities. For instance, during credential verification, a verifier uses a trust anchor to ensure the integrity of a subject's claims. Similarly, a user depends on a trust anchor to authenticate the source of their digital wallet software.

\subsubsection{Trust Anchor Model}
A solution can use one or more trust anchors, each serving different or overlapping purposes. For instance, a system might trust cloud providers for operations and rely on governmental agencies for credential issuance. Additionally, having multiple trust anchors can enhance assurance during audits by verifying trusted history and credential use.

\subsubsection{Trust Anchor Realization}
A trust anchor can be established technically or non-technically. For instance, a trust anchor for the tamper-proof distribution of cryptographic material across administrative domains can be implemented as a distributed ledger that is commonly operated and governed by all parties following a consortia agreement. In another approach, the government acts as a single trust anchor, dictating by regulation which entities are trustworthy for verifying specific identity attributes.

\textbf{Dimension Group 2: Identity:}
This group organizes identity-related dimensions and is divided into three subgroups: general identity dimensions, identifier-, and credential-specific dimensions.

\subsubsection{Governance Structure}
In this taxonomy, "Identity Governance" refers to where decision-making and authority of the non-ledger parts reside; whether in a single entity or distributed across multiple entities. Governance determines who can make and enforce rules about identity management processes, such as who can issue credentials, how disputes are resolved, and how system changes are approved.

\subsubsection{Subject}
This dimension specifies the types of entities that can possess identities within the IDM system. This dimension defines the scope and applicability of the identity solution. 

\subsubsection{Migration}
The migration of an identity focuses on how the identifiers and/or credentials can be transferred from one system or environment to another. The ability to migrate an identity is crucial for ensuring portability and interoperability.

\subsubsection{Lifespan}
The lifespan specifies how long an identity artifact remains valid following its creation. This can be permanent or limited to specific uses, such as with an ephemeral credential.

\begin{table*}
    \caption{The artifact "Taxonomy", showing DLT-based identity management architecture-related dimensions and their characteristics. The dimensions are grouped into three groups, which partially contain further subgroups.\\
    \footnotesize{ 
\textbf{Mutually Exclusive (ME):} [Y] = Within a given dimension, an object cannot simultaneously possess more than one characteristic. [N] = multiple characteristics can exist concurrently. 
}}
    \label{tab:taxonomy-v2-sandro}       
    \begin{tblr}{
        colspec = {|[1.5pt] c |[0.00pt] c | [1.5pt] X[0.20,c] |[1.5pt] c | [1.5pt] X[c] | [1.5pt]},
        hlines, 
        column{5-5}={colsep=10pt},
        column{3-3}={colsep=10pt}, 
        column{2-2}={colsep=0pt}, 
        column{1-1}={colsep=0pt}, 
        rows = {m},                
        row{1} = {font=\bfseries}, 
        rowsep = 1.5pt,           
        row{1} = {rowsep=6pt},
        row{2-24} = {rowsep=2pt}, 
        row{2-2} = {bg=gray!10},
        row{4-4} = {bg=gray!10},
        row{6-6} = {bg=gray!10},
        row{8-8} = {bg=gray!10},
        row{10-10} = {bg=gray!10},
        row{12-12} = {bg=gray!10},
        row{14-14} = {bg=gray!10},
        row{16-16} = {bg=gray!10},
        row{18-18} = {bg=gray!10},
        row{20-20} = {bg=gray!10},
        row{22-22} = {bg=gray!10},
        row{24-24} = {bg=gray!10},
        hline{1} = {1-6}{1.5pt},   
        hline{2} = {1-6}{1.5pt},   
        hline{5} = {1-6}{1.5pt},  
        hline{21} = {1-6}{1.5pt},  
        hline{24} = {1-6}{1.5pt},  
    }
    \SetCell[c=1]{c}\textbf{\ Group \&\ } &
    \SetCell[c=1]{c}\textbf{Subgroup\ \ } &
    \SetCell[c=1]{c}\textbf{Dimension} &
    \SetCell[c=1]{c}\textbf{ME} &
    \SetCell[c=1]{c}\textbf{Characteristics} \\

    \SetCell[r=3]{c,bg=white}{\rotatebox{90}{\parbox{3cm}{\centering \small \textbf{Trust Anchor}}}} &

    \SetCell[r=3]{c,bg=white}{\rotatebox{90}{\parbox{3cm}{\centering \small \textbf{}}}} &

     \SetCell[r=1]{c}{\textbf{Purpose}} & N &
     Trusted Issuer List \textbullet\  Trusted Verification Material \textbullet\ Trusted History of Verification Material \textbullet\ Trusted Operation \textbullet\  Trusted Storage of Credentials \textbullet\ Trusted Credential Status \textbullet\ Trusted History of Credential Status \textbullet\ Trusted History of Credentials Use \\
    
    &&
    \SetCell[r=1]{c}{\textbf{Model}} & N &
     Single Trust Anchor \textbullet\  Multiple Independent Trust Anchors \textbullet\  Multiple~Dependent~Trust~Anchors \\

    & &
     \SetCell[r=1]{c}{\textbf{Realization}} & N &
     Distributed Ledger Technology \textbullet\  Distributed Hash Table \textbullet\  Distributed File System \textbullet\ Hardware Security Architecture (e.g. TEE`s) \textbullet\  Certificate Transparency Logs \textbullet\ Web of Trust \textbullet\ Website \textbullet\ Contractual~Agreement \textbullet\ Regulation/Law \\

    \SetCell[r=16]{c,bg=white}{\rotatebox{90}{\parbox{2.2cm}{\centering \small \textbf{Identity}}}} &

    \SetCell[r=4]{c,bg=white}{\rotatebox{90}{\parbox{2.2cm}{\centering \small\textbf{General}}}} &

    \SetCell[r=1]{c}{\textbf{Governance}} & Y &
    Decentralized Governance \textbullet\ Consortium-based Governance \textbullet\ Central~Authority \textbullet\ By Regulation \textbullet\ No Governance  \\

    &&
    \SetCell[r=1]{c}{\textbf{Subject}} & N &
    Humans \textbullet\ Animals \textbullet\ Organizations \textbullet\ Devices \textbullet\ Smart~Contracts \textbullet\ Software~Applications~\textbullet\ Digital~Assets \textbullet\ Work~Loads \textbullet\ Other \\
    
    &&
    \SetCell[r=1]{c}{\textbf{Migration}} & Y &
    Without Issuer Interaction \textbullet\ With Issuer Interaction \textbullet\ Non-Transferable \\

    &&
    \SetCell[r=1]{c}{\textbf{Lifespan}} & Y &
    Permanent \textbullet\ Time-Limited \textbullet\ Activity-Based \textbullet\ Ephemeral \\

    &
    \SetCell[r=6]{c,bg=white}{\rotatebox{90}{\parbox{2.2cm}{\centering \small \textbf{Identifier}}}}

    &
    \SetCell[r=1]{c}{\textbf{Type}} & Y &
    Logical \textbullet\ Physical (Hardware-Bound) \textbullet\ Public-Key-Based \textbullet\ W3C~Decentralized~Identifier \textbullet\ Custom  \\

    &&
    \SetCell[r=1]{c}{\textbf{Anchored}} & Y &
    On-Chain Anchored \textbullet\ On-Chain Hashed \textbullet\ On Website \textbullet\ Self-Anchored \textbullet\ Not Anchored  \\

    &&
    \SetCell[r=1]{c}{\textbf{Proof Material}} & Y &
    On-Chain \textbullet\ Off-Chain Centralized \textbullet\ Off-Chain Decentralized \textbullet\ Off-Chain at Subject \textbullet\ Decentralized File System \\

    &&
    \SetCell[r=1]{c}{\textbf{Revocation}} &Y &
    No Revocation \textbullet\ On-Chain Revocation \textbullet\ Off-Chain Revocation \\

    &&
    \SetCell[r=1]{c}{\textbf{Recovery Mechanism}} & N &
    Multi-Signature Recovery \textbullet\ Time-Locked~Recovery \textbullet\ Secret~Sharing~\textbullet\ Escrow-Based~Recovery \textbullet\ Biometric~Recovery \textbullet\ Other~Mechanism \textbullet\ Other~Decentralized~Recovery \textbullet\ No~Recovery~Mechanism \\

    &&
    \SetCell[r=1]{c}{\textbf{Privacy}} & Y &
    Anonymous \textbullet\ Pseudoanonymous \textbullet\ Non anonymous \\

    &
    \SetCell[r=6]{c,bg=white}{\rotatebox{90}{\parbox{2.2cm}{\centering \small \textbf{Credential}}}}

    &
    \SetCell[r=1]{c}{\textbf{Type}} & N &
    Non-fungible Token \textbullet\ W3C~Verifiable~Credential \textbullet\ Other~Token~Standard \textbullet\  Smart-Contract-Based \textbullet\ Anonymous~Credential \textbullet\ Certificate \textbullet\ Custom  \\

    &&
    \SetCell[r=1]{c}{\textbf{Anchored}} & Y &
     On-Chain Credential Identifier \textbullet\ On-Chain~Hash \textbullet\ On-Chain~Encrypted \textbullet\ Not~anchored\\

    &&
    \SetCell[r=1]{c}{\textbf{Storage}} & Y &
    Off-Chain at Subject \textbullet\ On-Chain Plain Text \textbullet\ On-Chain Encrypted \textbullet\ Off-Chain Centralized \textbullet\ Off-Chain~Decentralized \textbullet\ Distributed~File~System \\

    &&
    \SetCell[r=1]{c}{\textbf{Revocation}} & Y &
    No Revocation \textbullet\ On-Chain Revocation \textbullet\ Off-Chain Revocation \\

   

    &&
    \SetCell[r=1]{c}{\textbf{Disclosure Control}} & Y &
    Only Full Disclosure \textbullet\ Combined Disclosure \textbullet\ Selective~Disclosure~\textbullet~Zero~Knowledge~Proof \\

    &&
    \SetCell[r=1]{c}{\textbf{Verifiability}} & Y &
    Non-Verifiable \textbullet\ Verifiable With Issuer Involvement \textbullet\ Verifiable~Without~Issuer~Involvement \\

    \SetCell[r=3]{c,bg=white}{\rotatebox{90}{\parbox{1.5cm}{\centering \small \textbf{Ledger}}}} 

    &
    
    \SetCell[r=3]{c,bg=white}{\rotatebox{90}{\parbox{4cm}{\centering \small }}}
        

    &
    \SetCell[r=1]{c}{\textbf{Consensus Mechanism}} & Y &
    Proof-of-Work \textbullet\ Proof-of-Stake \textbullet\ Proof-of-Authority \textbullet\ Delegated Proof-of-Stake \textbullet\ Byzantine Fault Tolerance \textbullet\ DAG-based \textbullet\ Other \\  

    &&
    \SetCell[r=1]{c}{\textbf{Permission Level}} & Y &
    Public Permissionless \textbullet\ Public Permissioned \textbullet\ Private Permissioned  \\

    &&
    \SetCell[r=1]{c}{\textbf{Usage Overhead}} & Y &
     No Fees \textbullet\ Creation Fee Only \textbullet\ Usage Fee Only \textbullet\ Creation and Usage Fees \textbullet\ Time-based (Subscription) \textbullet\ Alternative Fee Models 

  
  
  
  
    

    \end{tblr}
\end{table*}

\textbf{Identity Subgroup: Identifier}
\subsubsection{Type}
The identifier type refers to the form and nature of identifiers within the system. A physical identifier is a tangible object, like an ID card, while a logical identifier is a digital representation, such as an email address or DID.

\subsubsection{Anchored}
This dimension refers to where the source of truth of the identifier is located. Identifiers can be anchored on-chain transparently, as seen with many non-private DID methods. Alternatively, only the hash of the identifier may be stored on-chain. An example of non-anchored identifiers is peer DIDs, which are private, public-key-based identifiers.

\subsubsection{Proof Material}
The proof material refers to the location and type of location where the possession and control of the identifier is demonstrated. The self-contained characteristics indicate that the proof material is inherently included within the identifier(e.g. public-key).

\subsubsection{Revocation}
Revocation determines whether the architecture allows for invalidating an identifier and specifies where the revocation process is conducted and stored.

\subsubsection{Recovery Mechanisms}
The recovery mechanisms outline how a user can regain access to an identifier if the authenticator is lost.

\subsubsection{Privacy}
Privacy explains to what level the identifier is relatable to a subject's personal information.

\textbf{Identity Subgroup: Credential}
\subsubsection{Type}
Credential type refers to the various technical formats and standards used to implement the credential. The credential type is mutually inclusive, meaning two or more characteristics can apply to the same object under consideration. One example of ME credential type is a W3C VC implemented via a smart contract.

\subsubsection{Anchored}
This dimension indicates the location of the credential's source of truth. For instance, a system might have an on-chain credential identifier linked to an off-chain credential. Alternatively, the credential could be fully encrypted on-chain.

\subsubsection{Storage}
This dimension outlines credential storage locations and storage methods, including plain text, encrypted, or zk-proofs. Locations can be on-chain, off-chain, or both, and may be decentralized, centralized, or hybrid. The data format is not specified.

\subsubsection{Revocation}
Revocation determines whether the architecture allows for invalidating a credential and specifies where the revocation process is conducted and where the revocation status is stored. This dimension does not consider who is authorized to conduct revocation but only whether it is supported.

\subsubsection{Disclosure Control}
This concept relates to the amount of identity information shared during verification and the degree of control the subject has when disclosing a credential or specific parts of it.

\subsubsection{Verifiability}
Verifiability refers to whether a credential can be confirmed by a verifier. Credentials that are self-issued are not verifiable, while those from trusted issuers allow for independent verification without needing the issuer's involvement.

\textbf{Dimension Group 3: Ledger}

\subsubsection{Consensus Mechanism}
The consensus algorithm is a fundamental component of any distributed ledger-based solution and affects properties like fault tolerance, latency, decentralization, and security.

\subsubsection{Usage Overhead}
This dimension refers to how the DLT-based IDM system handles fees related to operational aspects such as the creation, usage, updating, and revocation of identity artifacts. Ledger-specific fees, such as transaction costs, are not included.

\subsubsection{Permission Level}
This dimension defines who can access the ledger. Public and private indicate its visibility, while permissioned means only authorized users can submit transactions, whereas permissionless allows anyone. Governance and participation rules are not covered by this dimension.
\section{Evaluation}
\label{sec:evaluation}
This section evaluates the taxonomy by applying it to two DLT-based IDM research papers and assessing the fulfillment of ending conditions.

\renewcommand{\arraystretch}{0.8}
\begin{table*}[htbp]
    \caption{Application of the taxonomy for classifying two proposed DLT-based IDM architectures.\\}
    \label{tab:taxonomy-v2-classification}       
    \begin{tblr}{
        colspec = {|[1.5pt] c |[0.00pt] c | [1.5pt] X[50pt,c] |[1.5pt] X[c] | [1.5pt] X[c] | [1.5pt]},
        hlines, 
        column{4-5}={colsep=10pt},
        column{3-3}={colsep=10pt}, 
        column{2-2}={colsep=0pt}, 
        column{1-1}={colsep=0pt}, 
        rows = {m},                
        row{1} = {font=\bfseries}, 
        rowsep = 1.0pt,           
        row{1} = {rowsep=2pt},
        row{2-4} = {rowsep=2pt}, 
        row{5-24} = {rowsep=2pt}, 
        row{2-2} = {bg=gray!10},
        row{4-4} = {bg=gray!10},
        row{6-6} = {bg=gray!10},
        row{8-8} = {bg=gray!10},
        row{10-10} = {bg=gray!10},
        row{12-12} = {bg=gray!10},
        row{14-14} = {bg=gray!10},
        row{16-16} = {bg=gray!10},
        row{18-18} = {bg=gray!10},
        row{20-20} = {bg=gray!10},
        row{22-22} = {bg=gray!10},
        row{24-24} = {bg=gray!10},
        hline{1} = {1-6}{1.5pt},   
        hline{2} = {1-6}{1.5pt},   
        hline{5} = {1-6}{1.5pt},  
        hline{21} = {1-6}{1.5pt},  
        hline{24} = {1-6}{1.5pt},  
    }
    \SetCell[c=1]{c}\textbf{\ Group \&\ } &
    \SetCell[c=1]{c}\textbf{Subgroup\ \ } &
    \SetCell[c=1]{c}\textbf{Dimension} &
    \SetCell[c=1]{c}\textbf{Classification of Solution 1\cite{xiong_bdim_2023}} &
    \SetCell[c=1]{c}\textbf{Classification of Solution 2\cite{grabatin_reliability_2018}} \\

    \SetCell[r=3]{c,bg=white}{\rotatebox{90}{\parbox{0.5cm}{\centering \small \textbf{Trust}}}} &

    \SetCell[r=3]{c,bg=white}{\rotatebox{90}{}} &

     \SetCell[r=1]{c}{\textbf{Purpose}} & Trusted verification material & N/A \\
    
    &&
    \SetCell[r=1]{c}{\textbf{Model}} & N/A & N/A \\

    & &
     \SetCell[r=1]{c}{\textbf{Realization}} & Distributed Ledger Technology & N/A \\

    \SetCell[r=16]{c,bg=white}{\rotatebox{90}{\parbox{2.2cm}{\centering \small \textbf{Identity}}}} &

    \SetCell[r=4]{c,bg=white}{\rotatebox{90}{\parbox{2.2cm}{\centering \small \textbf{General}}}} &

    \SetCell[r=1]{c}{\textbf{Governance}} & Central authority & N/A \\

    &&
    \SetCell[r=1]{c}{\textbf{Subject}} & Devices (IoT) & Organizations  \\
    
    &&
    \SetCell[r=1]{c}{\textbf{Migration}} & N/A & N/A \\

    &&
    \SetCell[r=1]{c}{\textbf{Lifespan}} & Permanent & Activity-based \\

    &
    \SetCell[r=6]{c,bg=white}{\rotatebox{90}{\parbox{2.2cm}{\centering \small \textbf{Identifier}}}}

    &
    \SetCell[r=1]{c}{\textbf{Type}} & Public-key-based \& W3C decentralized identifier & Logical (domain names)  \\

    &&
    \SetCell[r=1]{c}{\textbf{Anchored}} & On-chain anchored & On website \\

    &&
    \SetCell[r=1]{c}{\textbf{Proof Material}} & On-chain & Off-chain at subject \\

    &&
    \SetCell[r=1]{c}{\textbf{Revocation}} & On-chain revocation & N/A \\

    &&
    \SetCell[r=1]{c}{\textbf{Recovery Mechanism}} & No recovery mechanism & N/A \\

    &&
    \SetCell[r=1]{c}{\textbf{Privacy}} & Pseudoanonymous & Pseudoanonymous/non-anonymous depending on configurations \\

    &
    \SetCell[r=6]{c,bg=white}{\rotatebox{90}{\parbox{2.2cm}{\centering \small \textbf{Credential}}}}

    &
    \SetCell[r=1]{c}{\textbf{Type}} & Smart-contract-based \& W3C verifiable credential & Certificate (X.509) \\

    &&
    \SetCell[r=1]{c}{\textbf{Anchored}} & On-chain hash & Not-anchored \\

    &&
    \SetCell[r=1]{c}{\textbf{Storage}} & Off-chain at subject & Off-chain at subject \\

    &&
    \SetCell[r=1]{c}{\textbf{Revocation}} & N/A & N/A \\

    &&
    \SetCell[r=1]{c}{\textbf{Disclosure Control}} & N/A & Only full disclosure \\

    &&
    \SetCell[r=1]{c}{\textbf{Verifiability}} & Verifiable without issuer involvement & Verifiable, issuer involvement not mentioned \\

    \SetCell[r=3]{c,bg=white}{\rotatebox{90}{\parbox{1.5cm}{\centering \small \textbf{Ledger}}}} 

    &
    
    \SetCell[r=3]{c,bg=white}{\rotatebox{90}{\parbox{4cm}{\centering \small }}}
        
    &
    \SetCell[r=1]{c}{\textbf{Consensus Mechanism}} & Proof-of-stake & Other - depending on configurations \\  

    &&
    \SetCell[r=1]{c}{\textbf{Permission Level}} & Public permissionless & Private permissioned \\

    &&
    \SetCell[r=1]{c}{\textbf{Usage Overhead}} & No fees & No fees
    
    \end{tblr}
\end{table*}

\subsection{First Evaluation}
The first article proposes an architecture for decentralized identity management for Internet of Things (IoT) devices~\cite{xiong_bdim_2023}, addressing scalability and security challenges in large-scale IoT systems using smart contracts for access control, trust management, and reputation evaluation.

The classification result is shown in \autoref{tab:taxonomy-v2-classification} (Solution 1). The authors address many relevant dimensions, especially within the ledger and identity groups, though credential disclosure control, revocation, and migration capabilities are not discussed. The solution utilizes W3C DIDs and VCs in combination with PKI and smart contracts with various on-chain anchors and off-chain hashes.

Had the authors used this taxonomy, they might have recognized their trust anchor architecture lacks description of the model—whether using single or multiple anchors and relationships. While the authors present a comprehensive architecture, our taxonomy could have guided them to explore additional crucial dimensions.

\subsection{Second Evaluation}
The paper "SAML Metadata Management with Distributed Ledger Technology" \cite{grabatin_reliability_2018} explores managing SAML federation metadata with distributed ledger technology. The solution focuses on organizations as identity subjects and relies on PKI/X.509 certificates, suggesting the need for root Certificate Authorities, though trust anchor establishment is not explicitly discussed.
The solution uses logical identifiers (domain names) anchored through the ACME protocol on websites. Its privacy model leverages Hyperledger Fabric's channel-based architecture to restrict transaction visibility among specific participants. The credential system uses X.509 certificates stored off-chain, with no explicit mention of revocation or recovery mechanisms. The ledger employs a permissioned, private model with configurable consensus and no associated fees.
Applying the taxonomy to this paper reveals shortcomings including the lack of discussion on trust anchor establishment, governance structure, and revocation mechanisms. Additionally, by separating Identity into Identifier and Credential, the taxonomy highlights the tight coupling between identifier, verification material, and credential in X.509 certificates.

\subsection{Classification Results}
The taxonomy effectively classified two distinct architectural solutions, showcasing its versatility. The evaluation uncovered specific gaps that might have gone unnoticed without a systematic framework. The three-group structure organized complex interrelationships well, while separating Identity into general, identifier, and credential dimensions offered valuable clarity.
The taxonomy successfully categorizes both approaches without modification, demonstrating comprehensive dimension coverage and robustness validated by expert review. It enhances understanding of complex architectures and provides practical utility for both analysis and design. By highlighting architectural decisions and potential improvements, it serves as an analytical tool for researchers and a design guide for practitioners developing DLT-based identity solutions. 

\section{Discussion}
\label{sec:discussion}
This research developed a comprehensive taxonomy for DLT-based identity management through a rigorous iterative process. Trust emerged as the central concern across various layers of DLT-based IDM architectures, reflected in the taxonomy's organization into three groups: trust anchor, identity (further subdivided into general, credential, and identifier dimensions), and ledger-related dimensions.
The developed taxonomy represents a comprehensive artifact that significantly aids researchers and practitioners in designing and evaluating DLT-based IDM solutions. It provides a structured framework for classifying existing solutions, uncovering their shortcomings, and identifying improvements. The detailed nature of the taxonomy enables a nuanced analysis of both simple and complex architectures.
During the second iteration of the taxonomy building, there was an active discussion on how to structure the research approach and whether the meta-characteristics should be further specialized. It was decided to maintain a comprehensive and holistic view of architecture for the purpose of creating a detailed taxonomy artifact. All experts identified and agreed that trust is a significant concern for all identity management architectures, particularly those incorporating a distributed ledger component. When the taxonomy was analyzed through the lens of "trust," it became evident that nearly every dimension is also related to trust. Consequently, the trust anchor dimension was chosen in a way that fits multiple layers and points of view. We encourage future researchers and practitioners to examine and report their solutions through the lenses of this taxonomy.
The length and the revision steps of the dimensions may affect understandability. However, it was a deliberate choice to provide detailed and holistic coverage. The taxonomy underwent multiple revisions and expert reviews and was tested against existing taxonomies in the field. Another limitation could be the rapid evolution of the research field. However, we view the taxonomy as a living artifact and structured the dimensions to ensure their relevance and adaptability in the future.
Future general research is needed to assess where large language models can effectively assist researchers in building the taxonomy. Additionally, creating a standard way to report taxonomy artifacts, using interactive tools or software that apply the taxonomy, could enhance its usability and accessibility. Researchers could expand the taxonomy by creating application-specific groups and dimensions. They can also use this extensive taxonomy to create more specialized taxonomies for their specific fields and subsequently expand upon them. Moreover, both researchers and practitioners could test and refine our trust anchor framework. This taxonomy could be applied to classify a wide range of popular DLT-based IDM solutions, enabling their categorization and grouping, identifying shortcomings, and facilitating comparison.
\section{Conclusion}
\label{sec:conclusion}
This research presents the first comprehensive taxonomy at the intersection of IDM and DLT. Organized into three main components—trust anchor, identity, and ledger—the taxonomy provides researchers and practitioners with a systematic framework to analyze, compare, and design DLT-based identity solutions.
The taxonomy's significant contribution lies in establishing a common language and classification system for a rapidly evolving field. By structuring the complex landscape of DLT-based identity architectures, it enables more effective communication between researchers and practitioners, facilitates the identification of research gaps, and supports the development of more robust and interoperable solutions.
As distributed ledger technology continues to transform IDM approaches, this taxonomy offers a foundation for standardization and improved architectural design. It provides conceptual clarity about fundamental components while remaining adaptable to emerging technologies and implementation patterns. Through this structured approach to understanding DLT-based identity architectures, the taxonomy supports both theoretical advancement and practical implementation in this critical domain.

\bibliographystyle{IEEEtran}          
\bibliography{references/SLR_References/5+6+7+8+9+10+11,references/zotero-references-short,references/SLR_References/4_SecurityPrivacyTrust,references/SLR_References/1_SSI_Implementations, references/SLR_References/2_AuthAndAC,references/SLR_References/3_SurveysAndReviews, references/SLR_References/4_DIDs} 

\end{document}